\documentclass[12pt]{article}
\usepackage{epsfig}

\begin{document}

\title{Comparison and Combination of CRF Catalogues}
\author{Julia Sokolova, Zinovy Malkin}
\date{Pulkovo Observatory, St. Petersburg, Russia}
\maketitle

\begin{abstract}
In 2007, a joint IERS/IVS Working Group has been established to
consider practical issues of creating the next ICRF generation,
ICRF-2. The goal of the WG is to seek after ways to improve the
existing ICRF. In this study we investigate a possibility of ICRF
improvement by means of using combined ICRF catalogue instead of a
catalogue computed in a single analysis centre, even though using
most advanced models and software. In this work, we present a new
version of Pulkovo combined catalogue of radio source positions
computed using the method proposed in \cite{SokMal07}.
Radio source catalogues that were submitted in 2007 in the
framework of the WG activity were used as input for mutual
comparison and combination. Four combined catalogues have been
calculated: Two first of them provide a stochastic improvement of
the ICRF, and last two of them allow us to account also for
systematic errors in the current ICRF version.
\end{abstract}

\vfill
\noindent \hrule width 0.4\textwidth
~\vskip 0.2ex
\noindent {\small 5th IVS General Meeting, St.~Petersburg, Russia, 3--6 March 2007}
\eject

\section{Introduction}
The celestial reference frame (CRF), as realized by a set of
coordinates for selected celestial objects, is widely used for
numerous astronomy, navigation, time and other measurements. The
CRF accuracy and stability are all-important for successful
solutions of all these tasks. After publishing of the first VLBI
radio source catalogue (RSCs), attempts were made to improve the
accuracy of radio-band CRF by means of constructing combined
catalogues, as it was customary for optical astronomy, where
fundamental catalogues served as an international standard for
astrometry and other measurements on the sky. Different methods
were used to obtain a combined RSC, e.g.
\cite{Walter89a,Walter89b,Yatskiv90,Kurianova93}
Also, up to 1995, IERS (International Earth Rotation Service, now
International Earth Rotation and Reference Systems Service) used
Derived combined RSC for maintenance of the IERS Celestial
Reference Frame. In 2007, a joint IERS/IVS Working Group has been
established to consider practical issues of creating the next ICRF
generation, ICRF-2. The goal of the WG is to seek after ways to
improve the existing ICRF. Large experience accrued by optical
astrometry over centuries shows that combining catalogues of the
star positions leads to better random and systematic accuracy than
individual catalogues. In this work, we present a new version of
Pulkovo combined catalogue of radio source positions computed
using the method proposed in \cite{SokMal07}.
Radio source catalogues that were submitted in 2007 in the framework
of the WG activity were used as input for mutual comparison and
combination. Four combined catalogues have been calculated:
\begin{enumerate}
    \item The first two catalogues (the one has been calculated using all input
catalogues and the second one using only catalogues obtained with
CALC/SOLVE software) provide a stochastic improvement of the ICRF
    \item and the last two of them one allow us to account also for
systematic errors in the current ICRF version.
\end{enumerate}
All computations have been done for the set of 194 ICRF defining
sources included in all input catalogues.

\section{Input Catalogues and Comparisons}
\label{sec-tables}

\begin{table}[htb!] 
\caption{Information about input catalogues}
\label{tab-html_compat}
\begin{center}     
\tabcolsep=3pt
\small
\begin{tabular}{|l|l|c|c|l|}
  \hline
  IVS Centre & Soft & Time span (mon/yr) & \# of delays  & \# of sources \\
  \hline
  AUS, Australia & OCCAM & 11/1979 - 04/2007 & 2647809 & 1515  (212 def) \\
  BKG, Germany & Calc/Solve & 01/1984 - 10/2007 & 5156489 & 1076  (212 def) \\
  DGFI, Germany & OCCAM & 01/1980 - 01/2005 & 3650771 & 686  (199 def) \\
  CGS, Italy & Calc/Solve & 08/1979 - 07/2006 & 4640972 & 625  (161 def) \\
  USNO, USA & Calc/Solve & 08/1979 - 05/2007 & 5238056 & 923  (212 def) \\
  GSFC, USA & Calc/Solve & 08/1979 - 03/2007 & 5510462 & 923  (212 def) \\
  MAO, Ukraine & SteelBreeze & 04/1980 - 05/2007 & 5194922 & 2541  (26 def) \\
  IAA, Russia & QUASAR & 08/1979 - 05/2007 & 5116010 & 907  212 def) \\
  \hline
\end{tabular}
\end{center}
\end{table}

\begin{figure}
\centering
\includegraphics[scale=0.5]{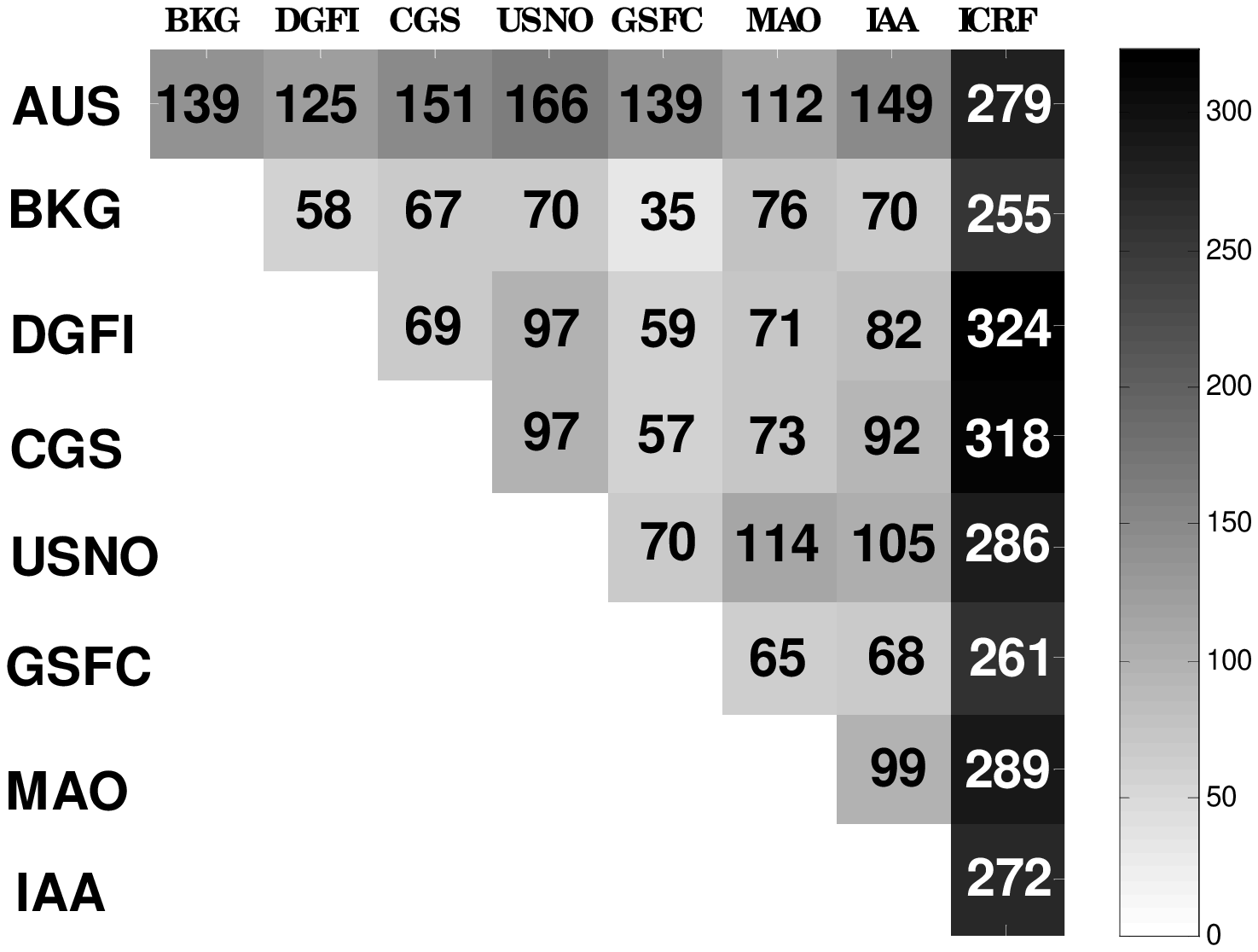}
\includegraphics[scale=0.5]{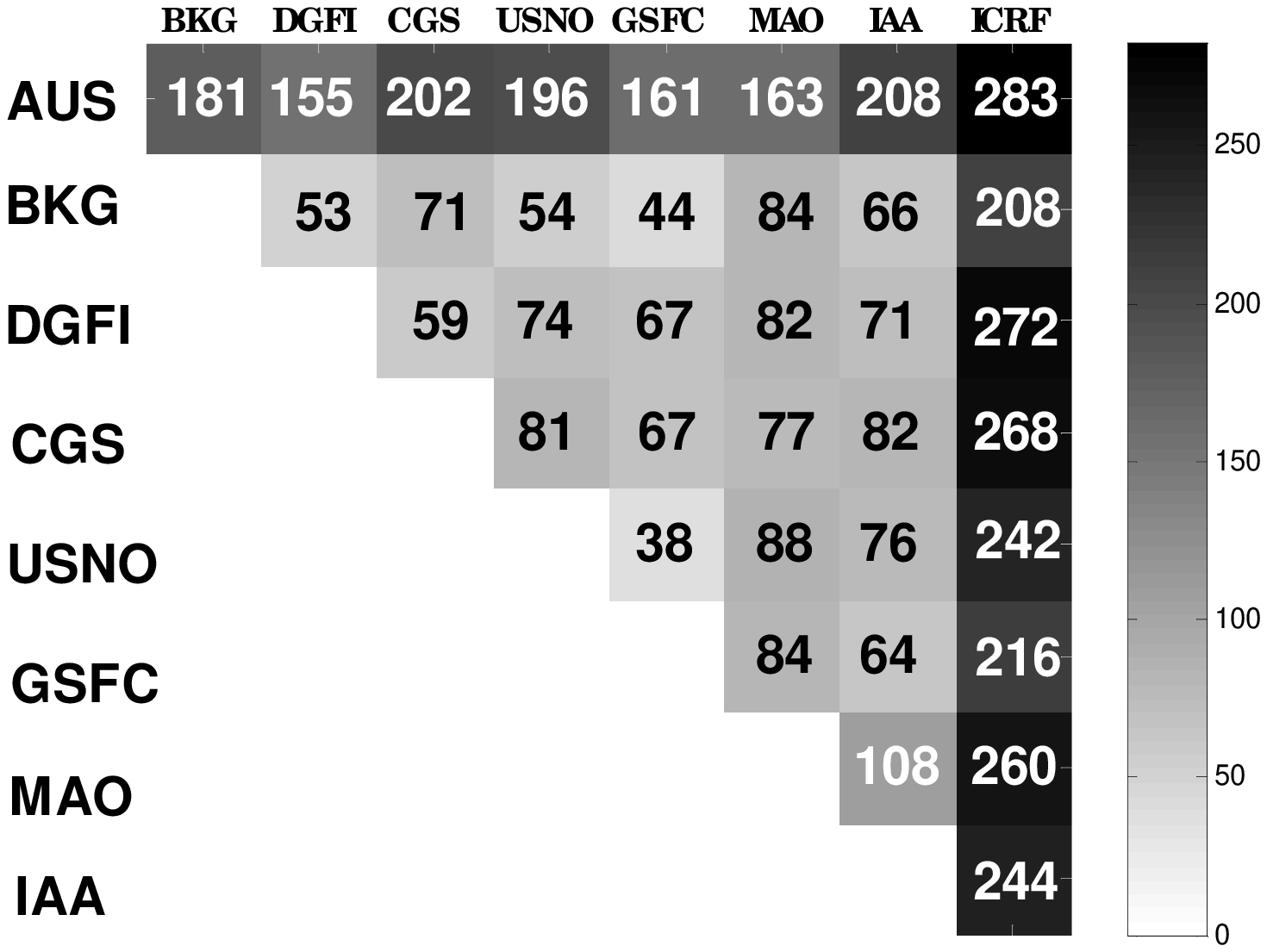}
\caption{WRMS of the intercomparison of the sources coordinates by
$\Delta\alpha$ (left) and by $\Delta\delta$ (right), For 194
common sources. Unit: Microarcseconds}
\end{figure}

Weighted root-mean-square (WRMS) differences of the radio source
coordinates between the input catalogues and ICRF are shown in
Fig. 1 (Unit microarcsecond). One can see from Fig. 1 WRMS
differences have the least values for catalogues computed with
Calc/Solve software, both for intercomparison of these catalogues
and their comparison with ICRF. The latter most probably is caused
by the fact that the ICRF was constructed using Calc/Solve
software. Moreover it can be clearly seen that all the input
catalogues demonstrate rather large differences with the ICRF,
which may indicate significant systematic errors in the latter.

\section{Combined catalogues}

The systematic differences between the input catalogues and the
ICRF found by the LF method were applied to all the input
catalogues in order to transform them to the ICRF system. After
that, the coordinates of all sources in transformed catalogues
were averaged with weights depending on the formal errors of
coordinates. In result, the combined catalogue PUL08C01a and
PUL08C01b (only input catalogues constructed using CALC/SOLVE were
used) were constructed.

\begin{figure}
\centering
\includegraphics[scale=0.5]{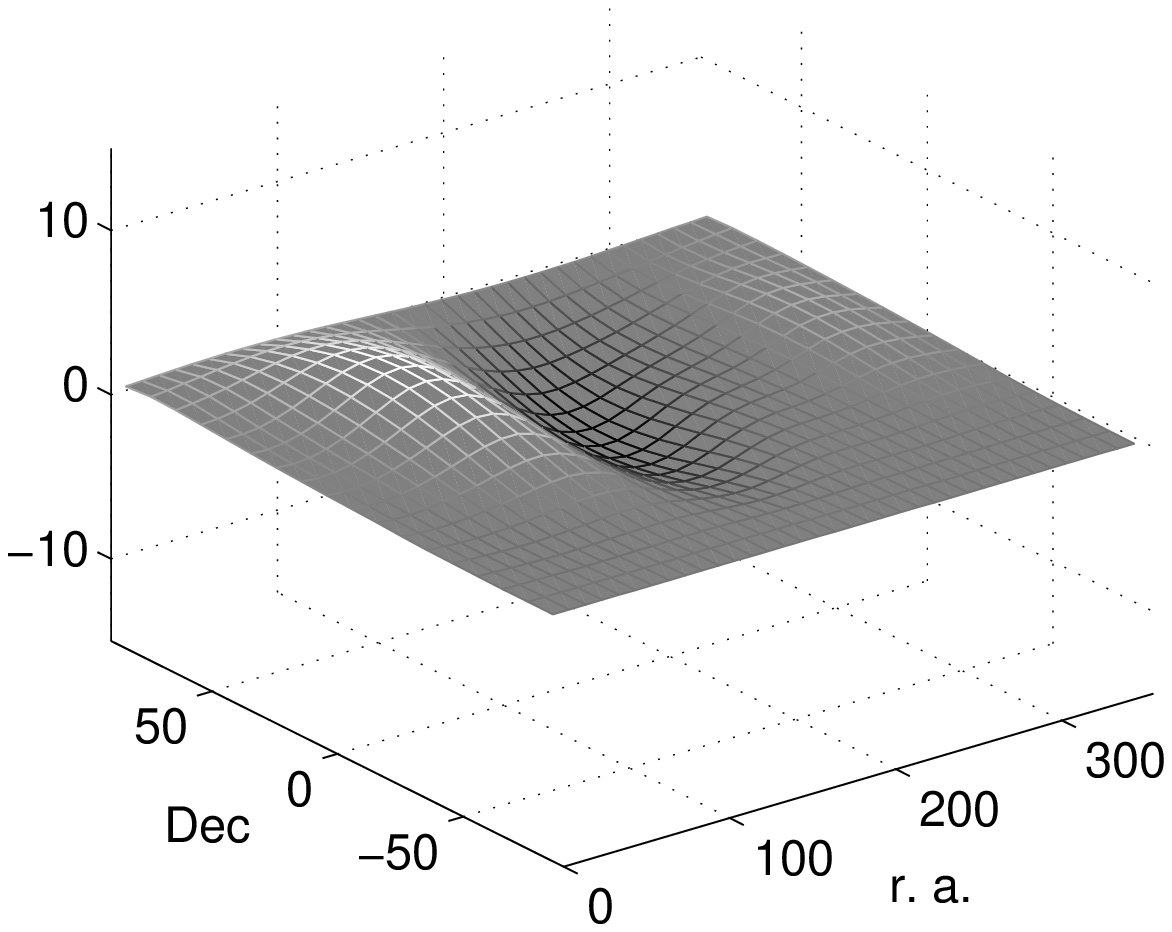}
\includegraphics[scale=0.5]{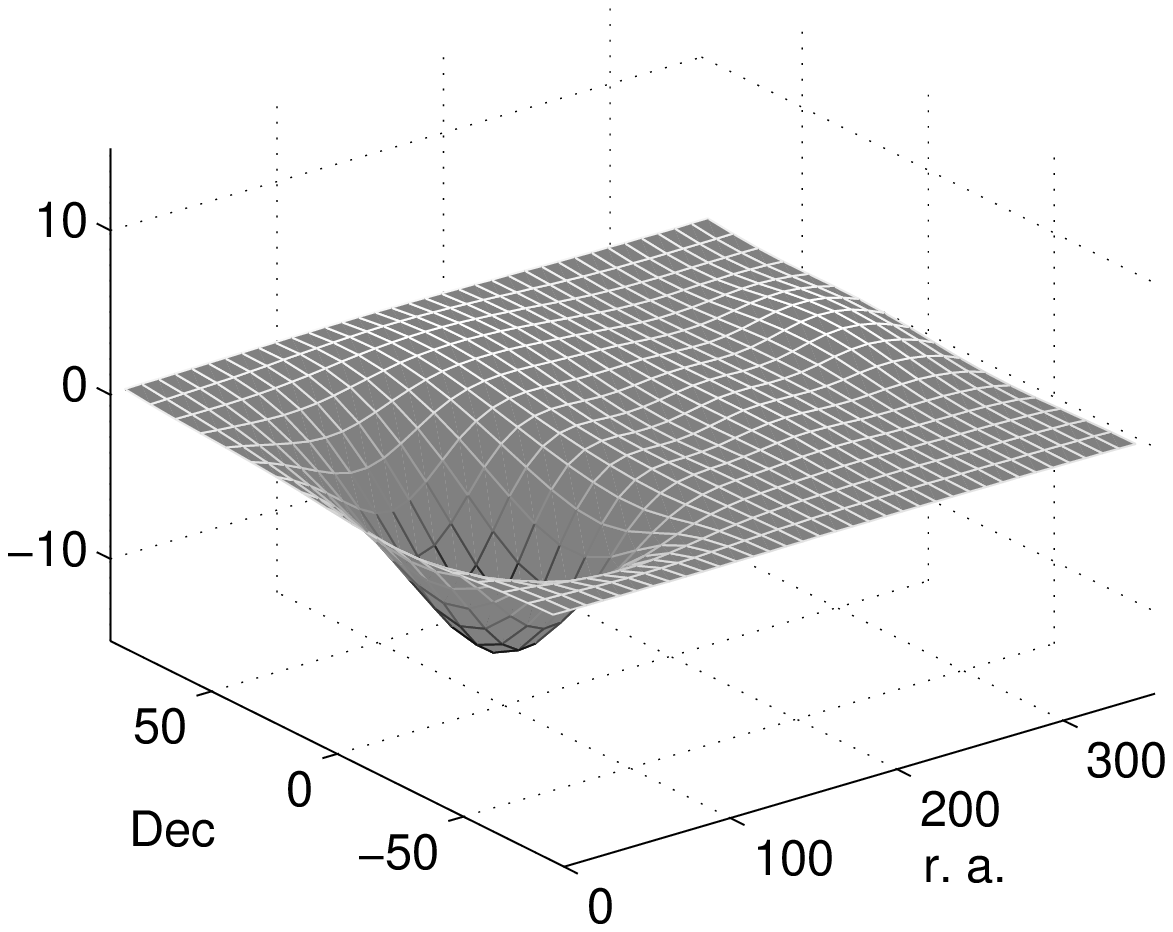}
\caption{PUL08C01a - ICRF, by $\Delta\alpha$ (left) and by $\Delta\delta$ (right), Unit: mas}
\end{figure}

\begin{figure}
\centering
\includegraphics[scale=0.5]{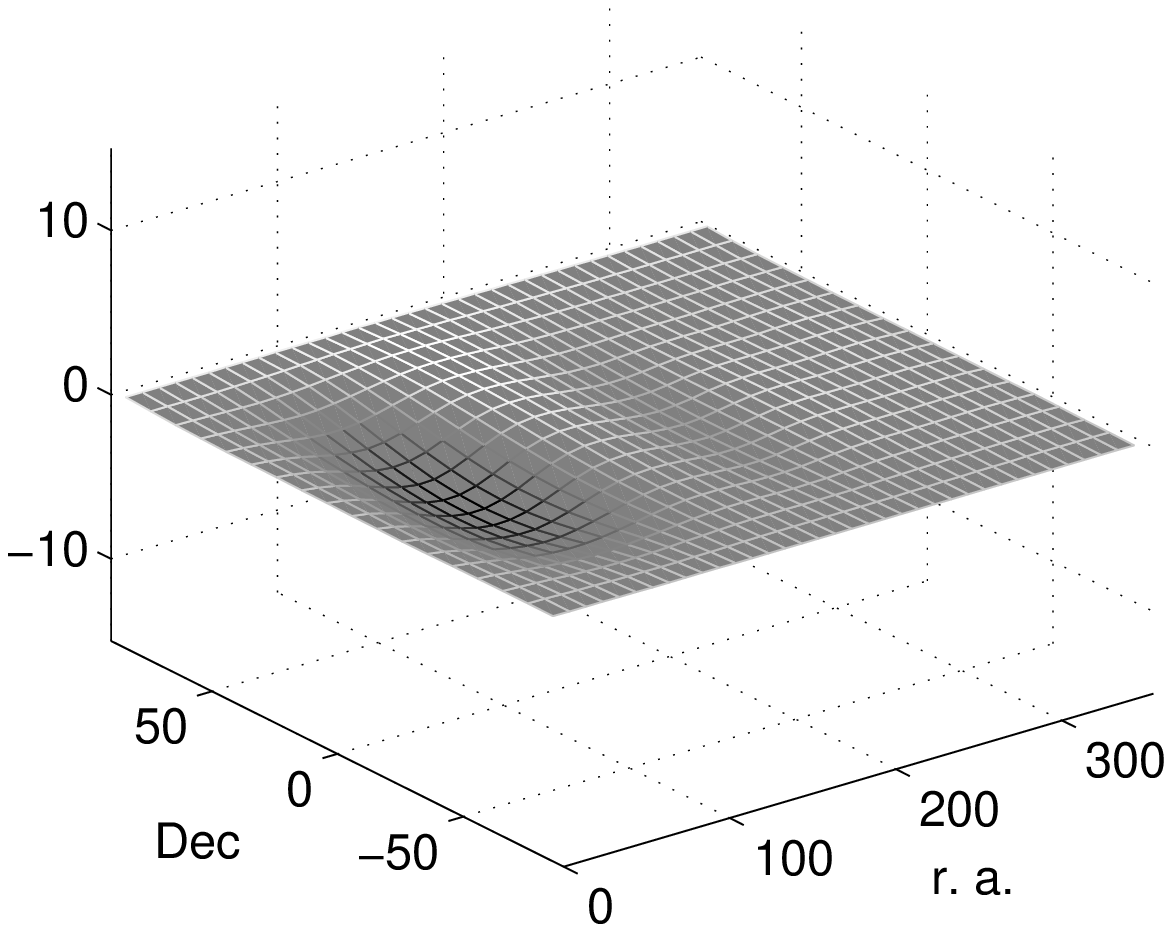}
\includegraphics[scale=0.5]{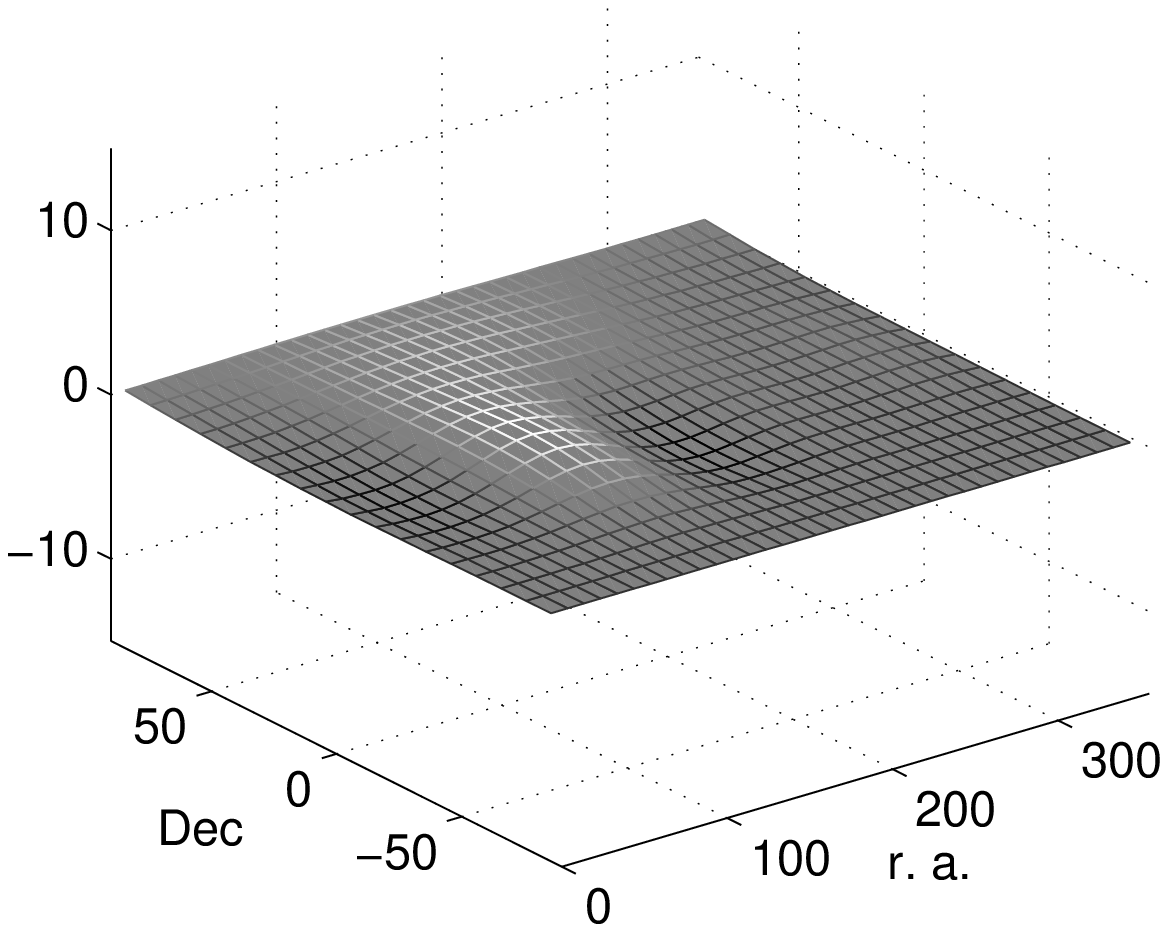}
\caption{PUL08C01b - ICRF, by $\Delta\alpha$ (left) and by $\Delta\delta$ (right), Unit: mas}
\end{figure}

Figure 2 show the systematic differences between the combined
catalogue PUL08C01a and the ICRF. One can see that the catalogue
PUL08C01a represents the ICRF system at a level of about 10
microarcseconds.

Figure 3 shows the systematic differences between the combined
catalogue PUL08C01b and the ICRF. The catalogue PUL08C01b
represents the ICRF system at a level of about 5 microarcsec.

\section{Short summary}
Four combined radio source catalogues have been constructed. The
first two of them, PUL08C01a and PUL08C01b can be considered as
stochastic improvement of the current realization of the ICRF. The
final combined catalogues PUL08C02a and PUL08C02b, provides both
stochastic and systematic improvement of the ICRF.

\end{document}